\documentclass[12pt,preprint]{aastex}
\usepackage{psfig,epsfig,amsfonts,amsmath,amssymb,graphicx,morefloats,appendix,tensor}
\usepackage{color}
\usepackage{natbib}
\usepackage{hyperref}
\begin{document}

\slugcomment{Accepted: March 23, 2021}

\title{Discovery of an Extremely Short Duration Flare from Proxima Centauri Using Millimeter through FUV Observations}

\author{Meredith A. MacGregor\altaffilmark{1}, Alycia J. Weinberger\altaffilmark{2}, R. O. Parke Loyd\altaffilmark{3},
Evgenya Shkolnik\altaffilmark{3}, Thomas Barclay\altaffilmark{4,5}, Ward S. Howard\altaffilmark{6}, Andrew Zic\altaffilmark{7,8}, Rachel A. Osten\altaffilmark{9,10}, Steven R. Cranmer\altaffilmark{1,12}, Adam F. Kowalski\altaffilmark{1,11}, Emil Lenc\altaffilmark{8}, Allison Youngblood\altaffilmark{12}, Anna Estes\altaffilmark{1}, David J. Wilner\altaffilmark{13}, Jan Forbrich\altaffilmark{14,13}, Anna Hughes\altaffilmark{15}, Nicholas M. Law\altaffilmark{6}, Tara Murphy\altaffilmark{7}, Aaron Boley\altaffilmark{15}, Jaymie Matthews\altaffilmark{15}}

\altaffiltext{1}{Department of Astrophysical and Planetary Sciences, University of Colorado, 2000 Colorado Avenue, Boulder, CO 80309, USA}
\altaffiltext{2}{Earth \& Planets Laboratory, Carnegie Institution for Science, Washington, DC 20015, USA}
\altaffiltext{3}{School of Earth and Space Exploration, Arizona State University, Tempe, AZ 85287, USA}
\altaffiltext{4}{NASA Goddard Space Flight Center, Greenbelt, MD 20771, USA}
\altaffiltext{5}{University of Maryland, Baltimore County, Baltimore, MD 21250, USA}
\altaffiltext{6}{Department of Physics and Astronomy, University of North Carolina at Chapel Hill, Chapel Hill, NC 27599, USA}
\altaffiltext{7}{Sydney Institute for Astronomy, School of Physics, University of Sydney, NSW 2006, Australia}
\altaffiltext{8}{CSIRO Astronomy and Space Science, Epping, NSW 1710, Australia}
\altaffiltext{9}{Space Telescope Science Institute, Baltimore, MD 21218 USA}
\altaffiltext{10}{Center for Astrophysical Sciences, Johns Hopkins University, Baltimore, MD 21218, USA}
\altaffiltext{11}{National Solar Observatory, University of Colorado Boulder, Boulder, CO 80303, USA}
\altaffiltext{12}{Laboratory for Atmospheric and Space Physics, University of Colorado, Boulder, CO 80303, USA}
\altaffiltext{13}{Center for Astrophysics \textbar~Harvard \& Smithsonian, Cambridge, MA 02138, USA}
\altaffiltext{14}{Centre for Astrophysics Research, University of Hertfordshire, AL10 9AB, UK}
\altaffiltext{15}{Department of Physics and Astronomy, University of British Columbia, Vancouver, BC V6T 1Z1, Canada}

\begin{abstract}

We present the discovery of an extreme flaring event from Proxima Cen by ASKAP, ALMA, HST, TESS, and the du Pont Telescope that occurred on 2019 May 1.   In the millimeter and FUV, this flare is the brightest ever detected, brightening by a factor of $>1000$ and $>14000$ as seen by ALMA and HST, respectively.  The millimeter and FUV continuum emission trace each other closely during the flare, suggesting that millimeter emission could serve as a proxy for FUV emission from stellar flares and become a powerful new tool to constrain the high-energy radiation environment of exoplanets.  Surprisingly, optical emission associated with the event peaks at a much lower level with a time delay.  The initial burst has an extremely short duration, lasting for $<10$~sec.  Taken together with the growing sample of millimeter M dwarf flares, this event suggests that millimeter emission is actually common during stellar flares and often originates from short burst-like events.   

\end{abstract}

\keywords{stars: individual (Proxima Centauri) --— 
stars: flare —-- 
stars: activity —--- 
submillimeter: planetary systems
}

\section{Introduction}
\label{sec:intro}

There has been extensive discussion of the prospects for life around low-mass, cool M-type stars, which are the most common stars in the Galaxy \citep{Henry:2006} and have a high frequency of Earth-sized planets with equilibrium temperatures allowing liquid water to be stable on their surfaces  \citep{Dressing:2015}. At the same time, M dwarfs exhibit higher levels of stellar activity and flaring throughout their entire lifetimes \cite[e.g.,][]{Schneider:2018,France:2016} compared to Sun-like stars. Over time, repeated large flares could deplete a planet's atmosphere of ozone themselves \cite[e.g.,][]{Tilley:2017} or due to associated energetic particles \cite[e.g.,][]{Segura:2010}, raising questions about the habitability of planets around these stars.  The Proxima Centauri system (Proxima Cen) is at the center of the habitability discussion because it is the closest exoplanetary system (1.3~pc) and has a potentially Earth-mass planet at a temperate $\sim$230~K equilibrium temperature \cite[semi-major axis $a\approx0.05$~AU,][]{AngladaEscude:2016}.  A second, more massive planet was recently discovered on a wider orbit \cite[$a\approx1.5$~AU,][]{Damasso:2020,Benedict:2020}.  Proxima Cen has long been known as a M-type flare star, making it a benchmark case for exploring the potential effects of activity \cite[e.g.,][]{Howard:2018} and strong stellar winds \cite[e.g.,][]{Garaffo:2016} on the planet's properties. Atacama Large Millimeter/submillimeter Array (ALMA) observations from 2017 resulted in the first observation of a M dwarf flare at millimeter wavelengths \citep{MacGregor:2018} opening a new observational window on the physics of stellar flares \citep{MacGregor:2020}.

\section{Survey Overview and Results}
\label{sec:survey}

We executed a multi-wavelength campaign to monitor Proxima Cen for $\sim40$~hours between April--July 2019 simultaneously at radio through X-ray wavelengths.  This paper presents the first results from this observing campaign, highlighting an extremely short duration flaring event observed on 2019 May 1 UTC by the Australian Square Kilometre Array Pathfinder (ASKAP), ALMA, the Transiting Exoplanet Survey Satellite (TESS), the du Pont telescope at Las Campanas, and the Hubble Space Telescope (HST). Details on the data reduction and analysis are provided in the Appendix.  Several other telescopes including Evryscope-South, The Las Cumbres Observatory Global Telescope (LCOGT) 1m, the Neil Gehrels Swift Observatory, and Chandra were involved in the full campaign but were not observing at the time of this event.  This observing campaign aligned with TESS observations in Sectors 11 and 12.  Several other analyses incorporating the available TESS data from this time period have been previously published by \cite{Vida:2019} and \cite{Zic:2020}.  However, the campaign presented here is unique in the multi-wavelength observations obtained simultaneously. Indeed, this is the first time that a stellar flare has been observed with such complete wavelength coverage (spanning millimeter to FUV wavelengths) and high time resolution (1~sec integrations with ALMA and HST) enabling unique insights into the process of flaring on M dwarfs. 

\begin{figure}[ht]
	\centering
	\includegraphics[scale=0.85]{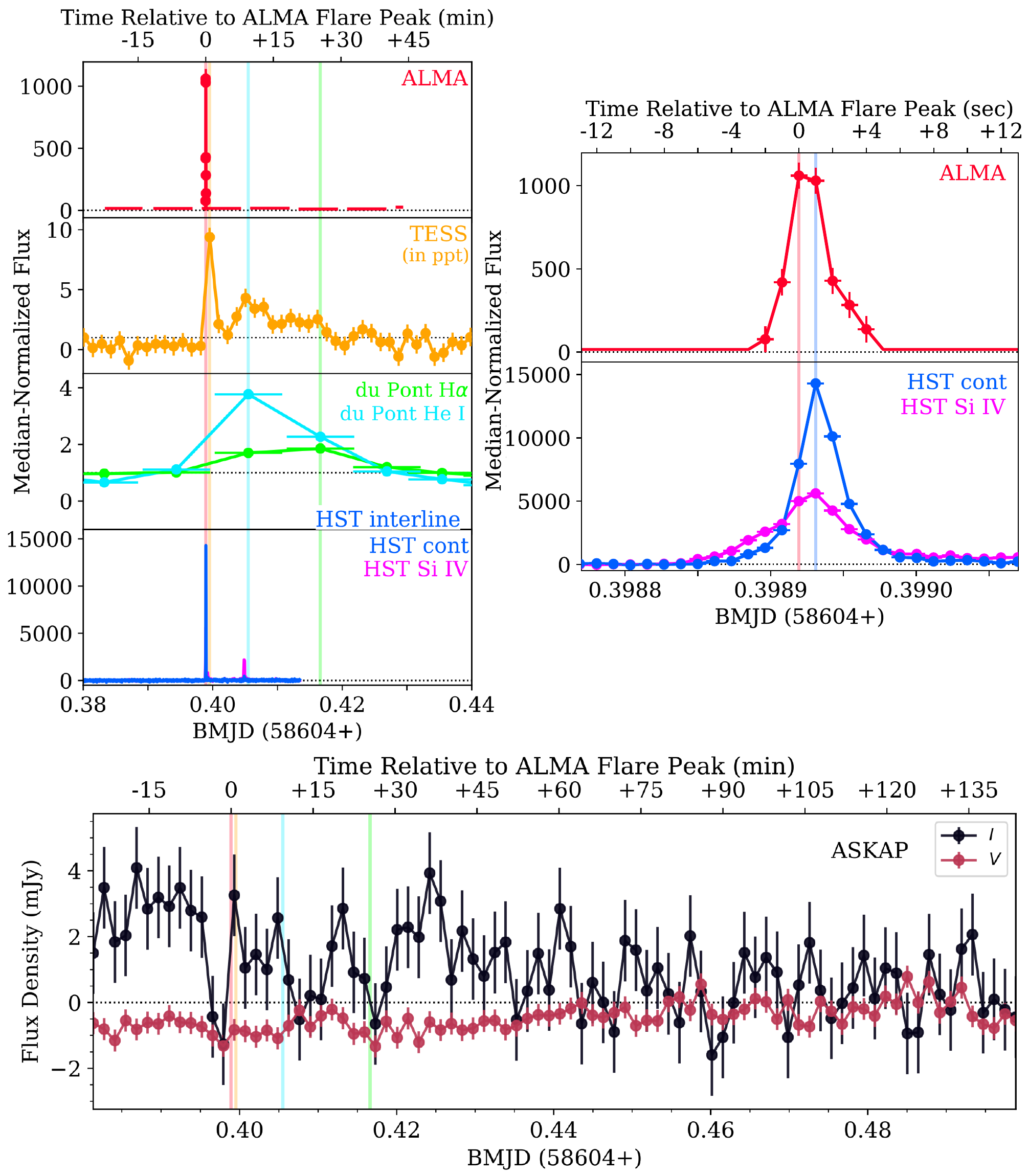}
	\caption{The complete light curve of the May 1 event at all wavelengths (left) and zoomed in to show the close correlation between millimeter and FUV as probed by ALMA and HST, respectively (right).  Median-normalized flux is plotted to compare the relative brightening during the flare compared to quiescence as observed by all facilities.  Vertical error bars indicate uncertainty in flux measurement, while horizontal error bars mark the time span of the individual observations from each telescope to aid the reader in interpreting potential emission delays between wavelength.  ASKAP Stokes I and V light curves for the entire night of observations surrounding the flare are shown at bottom.  The colored vertical lines (red--ALMA, orange--TESS, blue--du Pont He I, green--du Pont H$\alpha$, purple--HST continuum, magenta--HST Si IV) indicate the peak times of each facility to show correlation and delay.  ALMA does not detect detect quisescent emission from Proxima Cen, so the line plotted outside of the flaring event indicates a $3\sigma$ upper limit on the quiescent flux.}
	\label{fig:fig1}
\end{figure}

The complete light curve of the May 1 event at all wavelengths is shown in Figure~\ref{fig:fig1}.  In the millimeter and far-ultraviolet (FUV) pseudo-continuum, this is the brightest flare ever detected from Proxima Cen, brightening by a factor of $>1000$ and $>14000$ as seen by ALMA and HST, respectively.  Although the millimeter flare previously reported in \cite{MacGregor:2018} is nearly as bright, no counterparts were observed at any other wavelength making this new flare detection unique.  The corresponding optical signature observed by TESS brightens by a factor of only $0.9\%$, and peaks roughly 1~min later.  Although the precision of this delay is limited by the TESS 120-second integration time, the integration containing the optical flare peak does not overlap with the integration containing the millimeter and FUV flare peaks so some delay is indicated.  The event began as a strong, impulsive spike in the millimeter and FUV continuum with an initial rise time of $<5$~sec followed by a rapid drop on roughly the same timescale.  These properties have never been seen for a M dwarf flare before, suggesting that we could be observing an entirely new type of event.  The first ‘burst’ is followed $510$~sec later by a second smaller amplitude but longer duration event during which FUV line emission (e.g., Si IV) dominates with weak FUV continuum detected.  Many optical lines known as flare tracers that were not seen during the first peak also appear in visual-wavelength emission lines at the same time as the second event.  Unfortunately, this second event occurred during a calibration break and was not observed by ALMA.

\subsection{Radio and Millimeter Wavelengths}
\label{sec:radio}

The ASKAP observations show faint, $\sim50$\% circularly-polarized emission throughout the entire 14-hour observation, including a slowly-declining flux component that is not seen on any other day of the campaign. The emission has an average flux density of $1.15\pm0.14$~mJy and $-0.72\pm 0.09$~mJy in Stokes $I$ and $V$, respectively. We do not detect any radio burst counterpart. This apparent lack of correlation between low-frequency ($<1$~GHz) and higher frequency activity is commonly observed from active M-dwarf stars \citep{Kundu:1988}, and may indicate that the physical driver for low-frequency activity is independent of the processes driving flaring activity observed in higher-frequency~wavebands.

\begin{figure}[ht]
	\centering
	\includegraphics[scale=0.53]{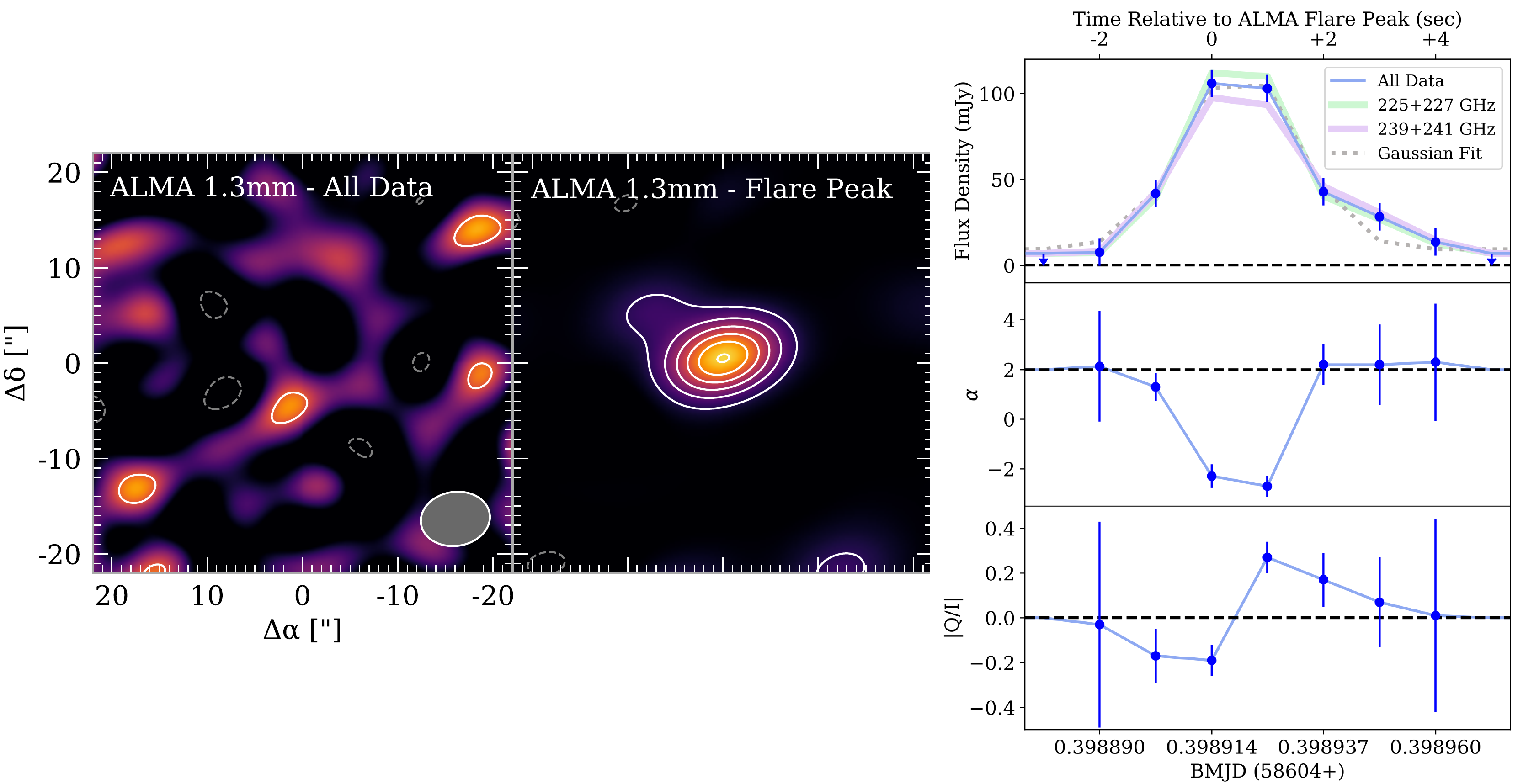}
	\caption{The $>1000\times$ brightness increased observed by ALMA during the flare is seen clearly by comparing an image of all ALMA data from May 1 (left) to the 1~sec integration at the flare peak (center).  Contours in both images are in steps of $2\times$ the rms noise of 80~$\mu$Jy and 7.9~mJy in the left and center images, respectively.  The ellipse in the lower right corner of the left panel shows the synthesized beam.  The spectral and polarization properties of the ALMA emission during this event are comparable to previously observed millimeter flares (right).  At peak (shown in the light curve in the top panel), the spectral index as a function of frequency (middle panel, $\alpha$, where $F_\nu \propto \nu^\alpha$) becomes steeply negative while the lower limit on the fractional linear polarization (bottom panel, $|Q/I|$) switches sign.  A simple Gaussian fit (shown by the gray dotted line in the top panel) reproduces the overall shape of the light curve.}
	\label{fig:fig2}
\end{figure}

As seen by ALMA (Figure~\ref{fig:fig2}, left), the May 1 flare reached a peak flux of $106\pm7.9$~mJy and luminosity of $2.14\pm0.15 \times10^{14}$~erg~s$^{-1}$~Hz$^{-1}$.  The right panel of Figure~\ref{fig:fig2} shows the ALMA light curve (top), along with the spectral index as a function of frequency ($\alpha$, middle), and a lower limit on the fractional linear polarization ($|Q/I|$, bottom).  This flare shows similar behavior to previous millimeter events \citep{MacGregor:2020}.  The light curve can be approximated by a Gaussian profile with a mostly symmetric rise and fall and no pronounced exponential tail.   The spectral index becomes steeply negative at peak, while the fractional polarization is initially negative before flipping positive during the short decay.  Outside of the flare and during the initial rise, the spectral index is consistent with emission from a quiescent stellar photosphere in the Rayleigh-Jeans regime.  The change in both the spectral and polarization properties of Proxima Cen during this event indicate a change in the dominant emission mechanism from thermal blackbody to synchrotron or gyrosynchrotron emission.

\pagebreak

\subsection{Optical Wavelengths}
\label{sec:optical}

\begin{figure}[ht]
	\centering
	{
		\includegraphics[trim= 9 5 9 5, clip, width=3.25in]{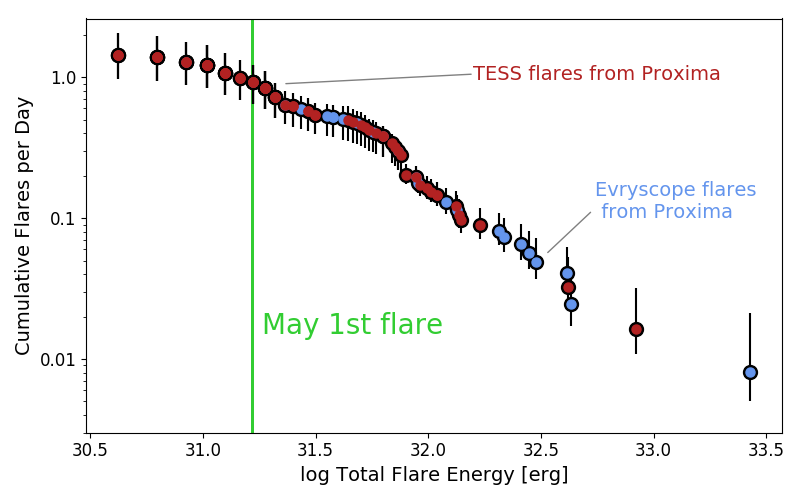}
		\label{fig:evr_vs_tess_proxima_flare_energy_FFD}
	}
	{
		\includegraphics[trim= 9 5 9 5, clip, width=3.25in]{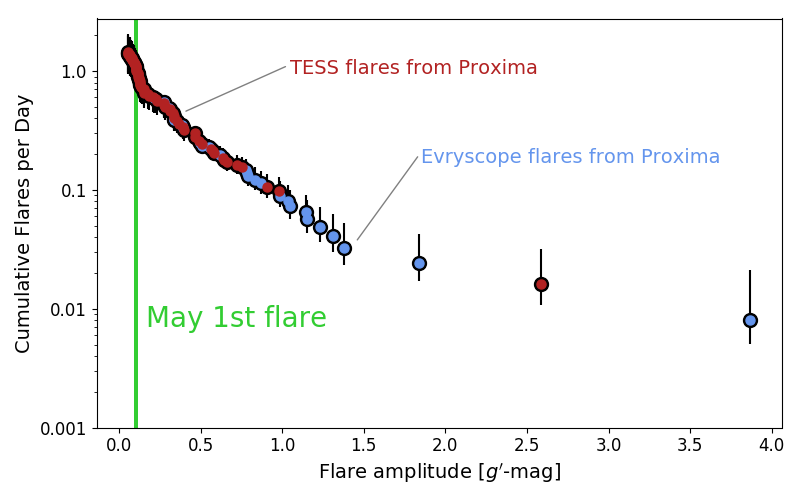}
		\label{fig:evr_vs_tess_proxima_flare_ampl_FFD}
	}
	\caption{These flare frequency diagrams (FFD) show that this flare is relatively weak in the optical when compared against all other stellar flares observed from Proxima by Evryscope and TESS. We find flares of equal or greater energy (left) and amplitude (right) to the May 1st flare occur once per day in the optical. Evryscope observations included span Jan 2016 to August 2018, while TESS observations span April 2019 to May 2019. Flare energies are converted from each bandpass to bolometric energy as described in Section~\ref{sec:optical}.}
	\label{fig:fig3}
\end{figure}

A bolometric energy of 10$^{31.2}$~erg was measured by \citet{Vida:2019} in TESS photometry for the initial May 1 event.  TESS observed 71 other flares from Proxima Cen with comparable energies of 10$^{30}$--10$^{32}$~erg across 50~days of observations \citep{Howard:2018}. Figure~\ref{fig:fig3} shows the flare frequency distribution (FFD) for the energy (left) and amplitude (right) of flares from Proxima Cen as observed by TESS and Evryscope-South. Larger flares make up $\sim$75\% of the combined sample \citep{Howard:2018}, and these FFDs predict flares with energies and amplitudes larger than the May 1 event occur once per day in the optical. The fact that the May 1 flare is apparently common in the optical yet extreme in the millimeter and FUV indicates that optical intensity does not necessarily scale to flare energies at other wavelengths and demonstrates the utility of multi-wavelength coverage.

Evryscope-South has monitored Proxima Cen with 2-minute cadence across 2~years of observations \citep{Ratzloff2019}. Since TESS observed Proxima for less than one stellar rotation near Proxima's activity minimum \citep{Wargelin:2017}, long-term flare monitoring with Evryscope provides a broader context to the flaring seen in the TESS data.  We convert optical flare energies from the TESS and Evryscope bandpasses into bolometric energies using a $\sim$9000 K blackbody. This canonical temperature provides an approximate fit to the spectrum of typical flares \citep{Osten2015}. We also convert the TESS flare amplitudes into the Evryscope $g^{\prime}$ bandpass using an $i$-to-$g$ scaling relation for flares from an M5.5 dwarf in \citet{Davenport:2012}. For the May 1 flare, these scaling relations yield an amplitude of 0.1 $g^{\prime}$ magnitudes.

We measure the cumulative flare frequency distribution (FFD) for the flare energy and amplitude by fitting a cumulative power-law to the Evryscope and TESS flares in Figure \ref{fig:fig3}. We calculate the uncertainty in the cumulative occurrence for each flare with a binomial $1\sigma$ confidence interval statistic, and estimate the uncertainty in our power-law fit through 1000 Monte-Carlo posterior draws consistent with our uncertainties in occurrence rates. We measure an FFD for the bolometric energies $E$ given by $\log{\nu_E} = \mathrm{-0.87}^{+0.15}_{-0.19} \log{E} + \mathrm{27.2}^{+6.2}_{-4.9}$, and an FFD for the $g^{\prime}$ amplitudes $A$ given by $\log{\nu_{A}} = \mathrm{-0.81}^{+0.10}_{-0.16} \log{A} + \mathrm{-0.05}^{+0.18}_{-0.15}$. $\nu$ is the number of flares per day of an equal or greater size to $E$ or $A$. Finally, because Evryscope has lower photometric precision than TESS, the smallest TESS flares are not always observed by Evryscope. We weight the flare rate of the smallest flares in the combined Evryscope and TESS FFDs by the TESS-only rates to remove bias from missing Evryscope flares. 

Figure~\ref{fig:fig4} (bottom left panel) shows the equivalent width curves over time for all of the emission lines we measured with du Pont during the May 1 flare. These emission lines display two types of behavior-- He~I, H$\beta$, H$\delta$, and H$\epsilon$ peak earlier, while Na D1 and D2, Ca H and K, H$\alpha$, and H$\gamma$ peak later.  Delayed Ca K emission has been observed previously and interpreted as evidence for chromospheric evaporation through the Neupert effect \citep{Kowalski:2013}.  The Balmer lines increase substantially in width during the flare.  H$\alpha$ is double peaked throughout the event, and the H$\beta$ line FWHM increases from $\sim$37 pre-flare to 45 km~s$^{-1}$ at peak.

\subsection{Far-Ultraviolet (FUV) Emission}
\label{sec:fuv}

\begin{figure}[t]
	\centering
	\includegraphics[scale=0.9]{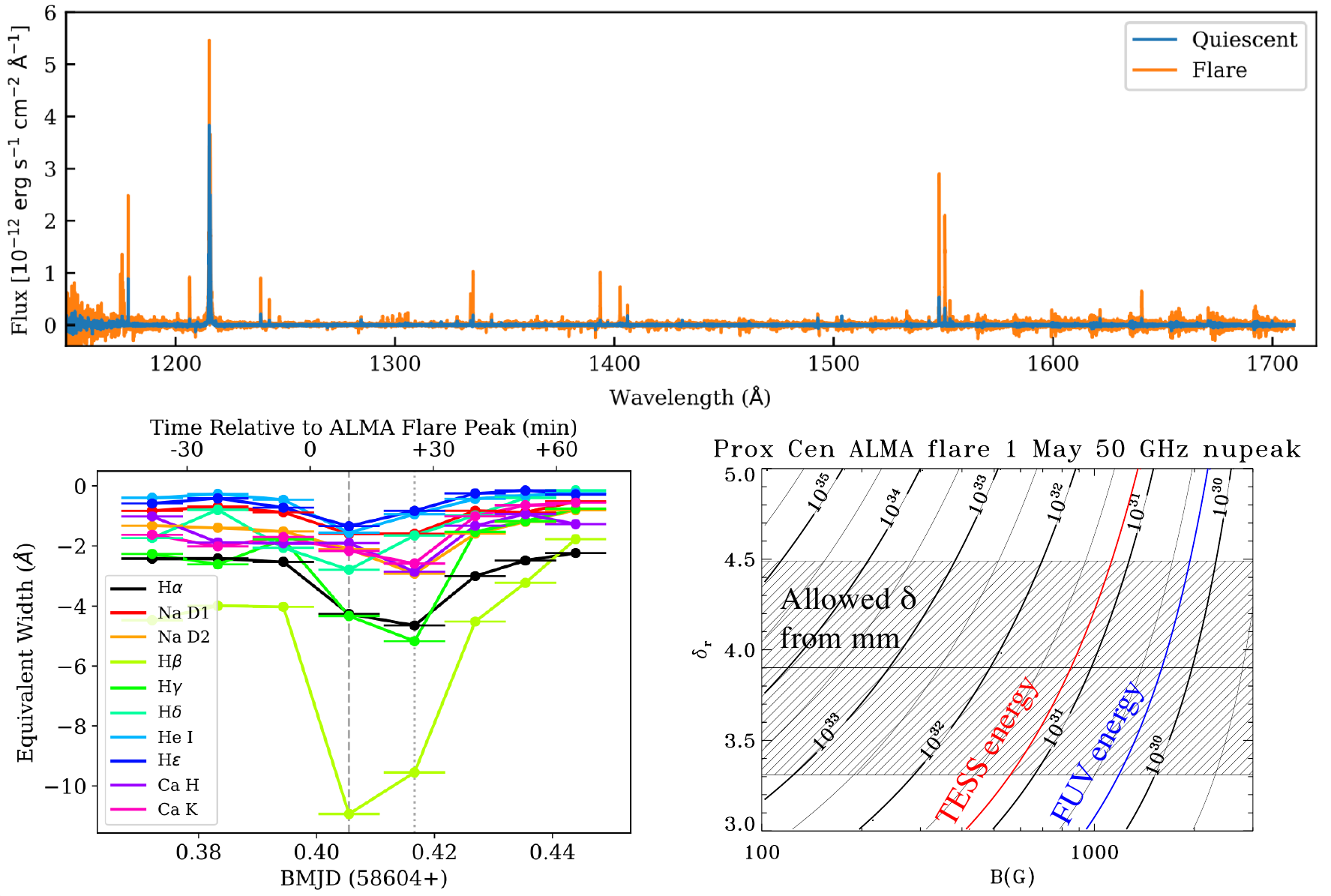}
	\caption{A comparison (top) between the HST spectra before (blue) and during the flare (orange) show that several strong emission lines appear only during the flare.  The equivalent width curves over time measured in the optical with du Pont (bottom left) show the dichotomy between lines that peak earlier and those that peak later, notably Ca K and H$\alpha$. The magnetic field strength in the radio-emitting source, $B$, can be constrained by the index of the accelerated particle distribution, $\delta_r$, and the nonthermal energy in accelerated particles (lower right).}
	\label{fig:fig4}
\end{figure}

The initial May 1 event exhibited an unprecedented rise in the star's pseudo-continuum FUV emission with an absolute FUV energy of $10^{30.3}$~erg, the largest ever recorded for Proxima Cen.  M dwarf flares reaching nearly $10^{33}$~erg have been observed previously by \textit{HST} from younger stars \citep{Loyd:2018a}, and the equivalent duration of quiescent emission required to produce the same energy as the flare, 4300~sec, is typical of daily UV flares on M dwarfs \citep{Loyd:2018b}. The color temperature of the pseudo-continuum flare emission is 15,000~--~22,000~K, covering 0.002~--~0.02\% of the visible stellar hemisphere (5$^\text{th}$~--~95$^\text{th}$ percentiles). A blackbody with these parameters would increase the flux in the \textit{TESS} bandpass by only 0.07~--~0.4\%, meaning additional mechanisms are required to explain the 0.9\% increase measured by \textit{TESS}. FUV line emission increases by a smaller factor of only $100-5000\times$ depending on the line, with broadening and asymmetric enhancements in redshifted emission out to $100$~km~s$^{-1}$ typical of stellar FUV flares \cite[e.g.,][]{Loyd:2018b,Hawley:2003}.  As seen in the top panel of Figure~\ref{fig:fig4}, several strong emission lines appear only during the flare, notably a singlet at 1247~\AA\ and a quintuplet centered on 1299~\AA\ that we have identified as originating from C$^{2+}$ and Si$^{2+}$ ions, respectively.  In the second flare, the pseudo-continuum reaches only 450$\times$ the pre-flare level while the lines increase 1000$\times$, which is more typical of FUV flares from M dwarfs \citep{Loyd:2018b}.

\section{Discussion}

The multi-wavelength coverage of this extreme event yields relative energies and timings, allowing us to examine correlations between flare emission at different wavelengths.  Only a few previous data sets include simultaneous FUV and optical coverage \cite[e.g.,][]{Hawley:2003}, and these have largely shown a correlation between flaring emission at these wavelengths.  The discrepancy between both the magnitude and timing of the optical emission compared with the millimeter and FUV emission that we see during the May 1 flare is new.  The correlation between the millimeter and FUV emission suggests that these wavelengths both directly trace the initial `impulsive' phase that releases most of the flare's energy as electrons accelerate in magnetic loops.  We interpret the emission at these wavelengths as arising from a hot blackbody with some contribution from synchrotron or gyrosynchrotron emission as indicated by the polarization signature at millimeter wavelengths.  The delayed optical emission likely comes from heated plasma at the loop footpoints. Delayed optical line emission, specifically H$\alpha$, is commonly observed in solar flares \cite[e.g.,][]{Benz:2017}, with H$\alpha$ emission peaking several minutes later during the `flash' phase, characterized by gentler energy release, possibly due to thermal conduction transport times \citep{Canfield:1990}.  The similar relative timing between wavelengths of the May 1 event sets-up a strong parallel to solar flares.

\subsection{Potential Emission Mechanisms}
\label{sec:emission}

The spectral and polarization properties at millimeter wavelengths suggest that we are seeing the optically thin part of the gyrosynchrotron spectrum.  We can infer the power-law index of non-thermal electrons, $\delta$, from the the spectral index, $\alpha$: $\alpha=1.22-0.9\delta_r$ \citep{Dulk:1985,Gudel:2002,Osten:2016}.  For the May 1 flare, this calculation yields $\delta_r = 3.9\pm0.59$, at the upper end of the range expected for hard radio spectra: $2.2\leq\delta_r\leq3.9$ \citep{Dulk:1985}.  Following the method in \citet{Osten:2016} and \citet{Smith:2005}, we assume a peak frequency of 50 GHz for the spectral energy distribution of gyrosynchrotron emission and calculate the integrated energy density in the flare over wavelength and time. This leads to a dependence of the total non-thermal energy on the index of the accelerated particle distribution, $\delta_r$, and magnetic field strength in the radio-emitting source. Contours corresponding to the bolometric flare energy (10$^{31.2}$ erg) and the FUV flare energy (10$^{30.3}$ erg) are in red and blue, respectively, in Figure~\ref{fig:fig4} (bottom right panel). A rough equipartition between non-thermal energy and the bolometric flare energy as observed by TESS (10$^{31.2}$ erg) constrains the magnetic field strength to be between about 400-1500 G. Larger field strengths occur if the non-thermal energy is closer to the FUV flare energy of 10$^{30.3}$ erg. Previous observations of Zeeman broadening in absorption lines of molecular FeH constrained the photospheric magnetic flux of Proxima Cen to be between $450-750$~G \citep{Reiners:2008}, consistent with our estimate from these new ALMA observations.  Interestingly, the spectral index of millimeter emission is the one characteristic of this event that differs significantly from solar flares.  While the May 1 flare and all previous millimeter M dwarf flares exhibit steeply negative spectral indices with frequency, solar flares at millimeter wavelengths typically have positive indices \citep{Krucker:2013}.

The extremely short duration of this event suggests that there is more to learn about the rapid time evolution of stellar and solar flares. \cite{Mouradian:1983} discussed the possibility that complex flares are composed of many elementary eruptive phenomena (EEPs).  Sequentially heated flare loops or `threads' form along an arcade structure, and combine or overlap to produce the total energy release of an event.  The $<10$~sec timescale of the initial May 1 burst roughly agrees with the duration of 10-20~sec X-ray bursts observed in short cadence solar observations \citep{Qiu:2012}.  Notably, previous flares observed by ALMA at millimeter wavelengths all have short timescales, ranging from $2-35$~sec \citep{MacGregor:2020}.  Perhaps millimeter emission is a common part of stellar flaring, and often originates from these short burst-like events not commonly seen at other wavelengths.

\subsection{Correlation Between Millimeter and FUV Emission}
\label{sec:mmfuv}

Measuring the FUV and EUV radiation environment of exoplanets is critical to predicting and
interpreting the chemical transformation and escape of their atmospheres \cite[e.g.,][]{Ranjan:2017,Owen:2019}.  However, observations at these wavelengths can only be performed from space and are limited by the dearth of facilities and absorption in the intervening interstellar medium.   These are the first simultaneous millimeter and FUV observations of a stellar flare and are therefore the first to show a strong correlation between emission at these wavelengths.  Notably, some previous observations have shown a correlation between microwave and hard X-ray emission during solar flares~\citep{Wiehl:1985}.  

If this trend is representative of flares in general, we can use the peak luminosity at millimeter wavelengths ($2.14\times10^{13}$~erg~s$^{-1}$~Hz$^{-1}$) and in the Si IV lines at 1393 and 1402~\AA~($5.41\times10^{27}$~erg~s$^{-1}$) to estimate the associated FUV luminosity for the millimeter flares previously observed with ALMA without simultaneous HST observations.  We accomplish this by deriving a scaling factor from the millimeter to Si IV (FUV) luminosity and applying it to previously millimeter measurements.  For the 2017 Proxima Cen flare \citep{MacGregor:2018}, the associated FUV luminosity might have been $5.1\times10^{27}$~erg~s$^{-1}$, while the previous largest AU Mic flare \citep{MacGregor:2020} might have had an FUV luminosity of $4.9\times10^{28}$~erg~s$^{-1}$.  This prediction is consistent with previous HST observations of flares from AU Mic, which had luminosities of $1.8-3.2\times10^{28}$~erg~s$^{-1}$ \citep{Loyd:2018b}.  It is striking that the FUV luminosity for the Proxima Cen and AU Mic events are within an order of magnitude of each other given the significant age and spectral type -- 4.85~Gyr, M5.5 \citep{Kervella:2003} and 22~Myr, M1 \citep{Mamajek:2014}, respectively -- difference between the two stars.  If millimeter emission can serve as a proxy for FUV emission from stellar flares, we will have a powerful new tool to determine stellar FUV emission, required input for models of planetary atmosphere evolution and abiotic oxygen accumulation \citep{Luger:2015}.

\section{Conclusions}
\label{sec:conclusions}

The observations of this one event observed by multiple facilities at millimeter to FUV wavelengths already challenges our current theoretical models of stellar flaring.  Proxima Cen is a unique target given that it hosts a planet in the habitable zone but also produces anomalously powerful flares for its old age.  Can a planet truly be habitable in this environment?  It is clear that necessary pieces are missing from our current understanding of M dwarf flares in order to answer that question.  We expect to learn much more as we synthesize the available data from this project and from future flare campaigns.  This paper presents the results from just a few minutes of the available data.  Many other flaring events are detected simultaneously across multiple facilities (including ALMA and HST) during the full 40-hour campaign.  If the correlation between FUV and millimeter flaring emission holds, there is potential for future all-sky millimeter surveys \cite[e.g., the Atacama Cosmology Telescope,][]{Naess:2020} to be able to provide constraints on the high-energy radiation environment of exoplanet host stars and inform discussion of planetary habitability.

\vspace{1cm}
This paper makes use of the following ALMA data: ADS/JAO.ALMA \#2018.1.00470.S.  ALMA is a partnership of ESO (representing its member states), NSF (USA) and NINS (Japan), together with NRC (Canada) and NSC and ASIAA (Taiwan), in cooperation with the Republic of Chile. The Joint ALMA Observatory is operated by ESO, AUI/NRAO and NAOJ.  This research is based on observations made with the NASA/ESA Hubble Space Telescope obtained from the Space Telescope Science Institute, which is operated by the Association of Universities for Research in Astronomy, Inc., under NASA contract NAS 5–26555. These observations are associated with program 15651. This paper includes data collected by the TESS mission. Funding for the TESS mission is provided by the NASA Explorer Program. The du Pont Telescope observations benefited from the assistance of Nidia Morrell and the staff of Las Campanas Observatory. 

MAM acknowledges support for part of this research from a National Science Foundation Astronomy and Astrophysics Postdoctoral Fellowship under Award No. AST-1701406. MAM and AJW acknowledge support from NRAO Student Observing Support (SOS) grants SOSPA6-011 and SOSPA6-021. Support for HST program number 15651 was provided by NASA through a grant from the Space Telescope Science Institute, which is operated by the Association of Universities for Research in Astronomy, Incorporated, under NASA contract NAS5-26555. TB acknowledge support from the GSFC Sellers Exoplanet Environments Collaboration (SEEC), which is funded in part by the NASA Planetary Science Divisions Internal Scientist Funding Model.

\software{\texttt{CASA} \cite[v5.3.0 \& v5.11.0][]{McMullin:2007}, \texttt{IRAF} \citep{Tody:1986,Tody:1993}, \texttt{Lightkurve} \citep{lightkurve}, \texttt{astropy} \citep{astropy:2013,astropy:2018}, \texttt{astroquery} \citep{Ginsburg:2019}}


\bibliography{References.bib}

\pagebreak

\section{Appendix}

Details on how the data were obtained, reduced, and analyzed are provided below for all facilities involved in the multi-wavelength campaign.  The wavelength range and exposure times for each instrument are listed in Table~\ref{tab:allobs}.  All observing times are reported in UTC and Modified Barycentric Julian Date (MBJD).  We have chosen to apply a barycentric correction, which accounts for differences in the Earth's position with respect to the barycenter of the Solar System, due to the wide spread in location between the ground- and space-based facilities involved.

\begin{table}[h]
\centering
\caption{Details of Multi-Wavelength Observing Campaign \label{tab:allobs}}
\begin{tabular}{lccc}

\hline
\hline
Facility  & Center & Wavelength & Avg. Integration\\
 & Wavelength & Range & Time (s)\\
\hline
ASKAP & 33.8~cm & 29.0--40.3~cm & 10\\
ALMA & 1.29~mm & 1.24--1.34~mm & 1\\
TESS & 775~nm & 580--970~nm & 120   \\
du Pont & 6675~\AA & 3500--9850~\AA & 800\\
HST & 1435~\AA & 1160--1710~\AA & 1 \\

\hline
\end{tabular}
\end{table}

\subsection{ASKAP}
We observed Proxima Cen with the Australian Square Kilometre Array Pathfinder \cite[ASKAP,][]{McConnell:2016} on 2019 May 1 09:01 UTC (scheduling block 8604) using a single on-axis (boresight) beam. We took the observation at a central frequency of 888\,MHz (33.8~cm) with 288\,MHz bandwidth, 1\,MHz channels, and 10\,s integrations, lasting 14\,hours in total. The primary calibrator PKS~B1934$-$638 was observed for 31 minutes immediately after the Proxima Cen observation (scheduling block 8606). We used on-dish calibrators to calibrate the frequency-dependent $XY$-phase.

We processed the ASKAP observations using the Common Astronomy Software Applications (\texttt{CASA}) package version~{5.3.0} \citep{McMullin:2007}. We determined flux scale, bandpass, and polarization leakage calibration using the observation of PKS~B1934$-$638. We used basic flagging routines to remove radio-frequency interference which corrupted approximately 20\% of the data.
        
We imaged the Proxima Cen field using the task \texttt{tclean} with a Briggs weighting and robustness of 0.0, using $6000\times6000$ pixels to include the full primary beam and the secondary sidelobes. We deconvolved the image using the \texttt{mtmfs} algorithm with a cell size of $2\farcs5$ and scale sizes of 0, 5, 15, 50, and 150 pixels. These imaging parameters enabled us to account for bright, complex field sources present both within and beyond the primary beam. To model sources with non-flat spectra, we used multi-frequency synthesis with two Taylor terms. We excluded a $4\arcmin$ square region centered on Proxima Cen from deconvolution to allow modeling of field sources without removing temporal and spectral variability of Proxima Cen. We deconvolved to a residual of $3\,\text{mJy}\,\text{beam}\,^{-1}$ to minimize PSF side-lobe confusion at the location of Proxima Cen. After deconvolution, we subtracted the derived field model from the calibrated visibilities using the task \texttt{uvsub}, and vector-averaged visibilities across all baselines longer than 200\,m for each instrumental polarization. 

Using the vector-averaged visibilities, we formed dynamic spectra for the four Stokes parameters ($I$, $Q$, $U$, $V$) consistent with the IAU convention of polarization, following \cite{Zic:2019}. To improve signal-to-noise ratio in the dynamic spectra, we averaged over a factor of 24 in time and 16 in frequency, giving a dynamic spectrum resolution of 240\,sec and 16\,MHz. We determined the rms noise in the dynamic spectra by taking the standard deviation of the imaginary part of the visibilities for each Stokes parameter, finding $3.4$\,mJy, $1.3$\,mJy, $1.5$\,mJy, $1.0$\,mJy for Stokes $I$, $Q$, $U$, and $V$ respectively. We formed light curves by averaging the dynamic spectra across frequency. The typical rms uncertainty in the light curves for Stokes $I$, $Q$, $U$, $V$ is $0.89$\,mJy, $0.36$\,mJy, $0.38$\,mJy, and $0.23$\,mJy. The higher level of rms noise in Stokes I products arise from imperfect subtraction of field sources, leading to enhanced levels of sidelobe confusion at the location of Proxima Cen.

To validate our visibility-domain approach to constructing light-curves, we imaged the Proxima Cen field in Stokes $I$ and $V$ at 4-minute intervals, measuring the peak flux density at the location of Proxima Cen. We found that the resulting light-curves agreed well with those from the visibility-based approach.

\subsection{ALMA}

The Atacama Compact Array (ACA) of ALMA observed Proxima Cen from 04:08:17 to 10:13:04 UTC on 2019 May 1 in four scheduling blocks (SBs) that each lasted approximately 93~min.  During each SB, Proxima Cen was observed in 6.5~min integrations or ‘scans’ alternating with a phase calibrator, J1424-6807, for a total of roughly 49~min on-source.  The flare discussed in this paper occurred during the final SB.  There were 9 antennas in the ACA during these observations spanning baselines of 10--47~m.  Flux and bandpass calibration were performed using the bright quasar J1337-1257. 

For these observations, the correlator was set-up to maximize continuum sensitivity.  Four spectral windows were centered at 225, 227, 239, and 241~GHz, with a total bandwidth of 2~GHz each, for a combined bandwidth of 8~GHz.  Two linear polarizations (XX and YY) were also obtained.  The raw ALMA data were reduced in \texttt{CASA} version 5.1.1 \citep{McMullin:2007} using the ALMA pipeline.  Deconvolution and imaging was performed using the \texttt{clean} task in \texttt{CASA}.  

We can use the broad bandwidth and dual polarization of these ALMA observations to determine both the spectral index as a function of frequency ($\alpha$, where $F_\nu \propto \nu^\alpha$) and a lower limit on the fractional linear polarization ($|Q/I|$) during the large flare on May 1.  We note that since we do not constrain Stokes U with these observations, $|Q/I|$ is only a lower limit to the true linear polarization fraction $p^2_{QU} = (Q/I)^2 + (U/I)^2$.  In order to calculate the spectral index of the flaring emission, we fit independent point source models to the millimeter visibilities from both the lower and upper sidebands.  At peak, the flux densities are $112\pm10.4$~mJy and $97.6\pm11.0$~mJy for the lower and upper sidebands, respectively, yielding a spectral index of $\alpha = -2.29\pm0.48$.  We checked the accuracy of the frequency-dependent amplitude calibration performed by the ALMA pipeline by comparing the results for the flux, bandpass, and phase calibration sources to the ALMA calibrator catalogue \citep{Bonato:2018}.  To determine a lower limit on the fractional linear polarization, we fit point source models to the XX and YY visibilities independently and compute the Stokes $Q = \langle E_X^2 \rangle - \langle E_Y^2 \rangle$ and $I = \langle E_X^2 \rangle + \langle E_Y^2 \rangle$, where $E_X$ and $E_Y$ are antenna voltage patterns.  This process yields $|Q/I| = -0.19 \pm 0.07$ at flare peak.

\subsection{TESS}
TESS is equipped with four cameras that each cover one $24\times24$ degree field of view and have an approximate effective aperture size of 10 cm. The throughput of TESS is optimized for the red-optical, with $>$50\% throughput at 580--970 nm. TESS is in a high earth orbit, with an average orbital period of 13.7 days. The four TESS cameras are aligned to project onto the sky in a 1x4 grid and the field of view is adjusted every two orbits. TESS integrates continually using frame transfer CCDs, which are read out every 2~sec. The full $96\times24$ degree field of view is stored by the spacecraft every 30 min, while select targets are stored at a high cadence of 2-minutes. These 2-minute cadence observation represent an integration time of 96~sec, because 20\% of data is rejected by the on-board cosmic ray mitigation algorithm. Proxima Cen was observed by TESS at 2-minute cadence from 2019-Apr-22 to 2019-Jun-19, covering two observing sectors (Sectors 11 and 12). The May 1 flare occurred during Sector 11.

We analysed the TESS data starting with calibrated pixel-level data rather than calibrated light-curves because the TESS sytematic removal is not designed to handle impulsive outliers, such as flares. We used the \texttt{Lightkurve} software package \citep{lightkurve} and dependencies \texttt{astropy} \citep{astropy:2013,astropy:2018} and \texttt{astroquery} \citep{Ginsburg:2019} to download the data from the Mikulski Archive for Space Telescopes (MAST) archive, and to extract a light curve from the pixel-data. We used the `threshold' method to determine the pixel mask used to create the light curve, with the threshold set at 3-sigma. We removed instrumental systematics using the Pixel Level Decorrelation (PLD) algorithm \citep{Deming:2015,Luger:2016} implemented in \texttt{Lightkurve}.

We took care to explore the impact of including different pixels in both the light curve extraction aperture and in the PLD analysis region on the May 1 flare. While the May 1 flare was clearly detected by TESS, the exact shape of the flare is somewhat challenging to resolve because of the relatively low signal-to-noise detection of the flare and its short duration relative to the TESS sampling cadence. We found that the primary peak of the flare (this is the time period that overlaps with the ALMA flare) was insensitive to changes in the pixels used in the analysis but the flux in the shape of the decay phase of the flare, the second and third peaks in the TESS data, were susceptible to being over fit by the PLD algorithm. We chose a set of pixels so that the PLD model did not have strong variability during the flare, as this likely represents a solution that is not subject to overfitting. 

\subsection{du Pont Telescope}
The Echelle Spectrograph on the 2.5~m Ir\'en\'e du Pont Telescope at Las Campanas Observatory provides complete wavelength coverage from $\sim$3500 to 9850~\AA\ at a resolution of $\sim$40,000 in a $1\arcsec$ slit.  The weather was clear and seeing $\sim$0\farcs9. Exposures were taken starting at 04:30 UT and continuing until 10:39 UT with the only gaps caused by readout of the CCD. Exposure times ranged from 600-900~sec during the flare, with shorter exposures taken during the peak because the S/N on the emission lines was noticeably higher. ThAr lamp spectra were taken at the beginning and end of the sequence. 

\texttt{IRAF} \citep{Tody:1986,Tody:1993} tasks were used for the data reduction from the 2D CCD images through extraction and wavelength calibration. Basic CCD reduction began with overscan subtraction, flat fielding, and bad pixel correction.  The flatfield was made by averaging together 52 flats taken between 25 April and 06 May. Each flat is a daytime sky spectrum taken through a diffusing glass and then divided by a 15x15 pixel box-car filtered version of itself. Spectra were extracted with the {\it apall} package. For orders in which the continuum was traceable, each extraction aperture was recentered on each order. For orders at $\lambda<$4400\AA\ that had bright emission lines, but little continuum, the apertures were centered based on the emission line position. Background regions were set individually for each order. The extraction was variance weighted.  Wavelength calibration was based on only the ThAr lamp taken at the end of the night, i.e. at 10:48 UT.

Using a custom IDL code, we fit the continuum with a 3rd order polynomial and subtracted it. Also with a custom code, we measured the equivalent widths of the Hydrogen Balmer series (H$\alpha$ through H$\epsilon$), Ca H and K, He I, and Na I D1 and D2 lines.  Integration limits were set by visual inspection of each spectrum. The per pixel uncertainty was estimated from the noise in the continuum on either side of the line and then propagated into an uncertainty on the equivalent width.  

Six spectra cover the start of the flare through the end of the night (Table~\ref{tab:echellelog}). The spectrum containing the UV flare started exposing at 09:13:38 UT and had an exposure time of 900~sec. The first exposure with substantial flare emission started at 09:29:40 UT and also had an exposure time of 900~sec.

\begin{table}[h]
\centering
\caption{Observing Log for du Pont Echelle \label{tab:echellelog}}
\begin{tabular}{ccc}

\hline
\hline
Exposure Start &Exposure Middle &Exposure Time\\
(UT)                &(MJD - 2458604) &(sec)\\
\hline
09:13:38 &0.894369 &900\\
09:29:40 &0.905503 &900\\
09:45:41 &0.916626 &900   \\
10:01:43 &0.926893 &750\\
10:15:14 &0.935411 &600\\
10:26:16 &0.943942 &750 \\

\hline
\end{tabular}
\end{table}

\subsection{HST}
We collected the HST data using the Space Telescope Imaging Spectrograph (STIS) instrument with the E140M grating.  The data are in time-tag format, meaning STIS operated as a photon counting detector.  The spectral resolving power of the instrument is $R=\lambda/\Delta\lambda=45,800$ and time resolution is, in practice, limited by signal to noise rather than the instrumental resolution.  Greater time resolution is possible in spectral and temporal ranges with greater flux.  The intrinsic time-tagging of the MAMA detector  is 125~microseconds.

We obtained the reduced HST data products from the MAST archive on 2019 Nov 8.  We then used a custom code to calibrate the wavelength and apply effective-area weighting to photons in the raw time-tag (``\_tag.x1d'') data files.  The code extracted events from regions on the detector where the spectral traces of the Echelle fall, then used regions without signal above and below the spectral trace to estimate and subtract a background count rate.  In this way, it generated time-integrated spectra that it then compared to the fully pipeline-reduced spectra to estimate the photon wavelengths and effective-area weights, producing a calibrated photon list.  From the calibrated photon lists, we integrated photons with arbitrary wavelength and time binning to generate light curves of emission from specific wavelengths and spectra of emission integrated over intra-exposure time ranges.  This is the same process used by \cite{Loyd:2018a} and \cite{Loyd:2018b}.  Using our custom code avoided background oversubtraction issues that appeared when generating sub-exposure spectra with the HST \texttt{CALSTIS} tools.

\end{document}